\documentclass[aps,preprint]{revtex4}%
\usepackage{amsfonts}
\usepackage{amsmath}
\usepackage{amssymb}
\usepackage{graphicx}%
\providecommand{\U}[1]{\protect\rule{.1in}{.1in}}

\begin{document}
\title[ ]{Point-Charge Models and Averages for Electromagnetic Quantities Considered in
Two Relativistic Inertial Frames}
\author{Timothy H. Boyer}
\affiliation{Department of Physics, City College of the City University of New York, New
York, New York 10031}
\affiliation{E-mail: boyer@sci.ccny.cuny.edu}
\keywords{}
\pacs{}

\begin{abstract}
Electromagnetic quantities at a spacetime point have tensor Lorentz
transformations between relatively-moving inertial frames. \ However, since
the Lorentz transformation of time between inertial frames depends upon both
the time and space coordinates, \textit{averages} of electrodynamic quantities
at a single time will in general depend upon the inertial frame, and will
differ between inertial frames. \ Here we illustrate how the use of continuous
charge and current distributions rather than point-charge distributions can
lead to physically mystifying and even inaccurate results for electromagnetic
quantities and physical phenomena. \ The discrepancy noted between the average
electric field values in different inertial frames is particularly striking
because it is first order in the relatative velocity between the frames. \ 

\end{abstract}
\maketitle

\section{Introduction}

\subsection{Apparent Relativity Paradox}

There is an apparent relativity paradox involving magnets which has puzzled
some teachers of electromagnetism. A toroid is claimed to have negligible
electric and magnetic fields outside the surface currents. Thus, using tensor
transformations for the fields, we should expect that there should be
negligible electromagnetic fields outside the magnet in an inertial frame in
which the magnet is moving. However, there is a problem in Griffiths'
junior-level electromagnetism text\cite{G571} and in the first edition of
Jackson's graduate text\cite{Jackson} (and only the first edition) suggesting
that in an inertial frame in which the magnet is moving perpendicular to the
magnetic field inside its winding, the magnet develops a non-zero scalar
potential outside the magnet. The existence of a scalar potential suggests the
possibility of electric fields outside the magnet in the inertial frame in
which the magnet is moving. Indeed, the electric fields outside a magnet in
the inertial frame in which it is moving have been used in a classical
electromagnetic analysis\cite{B2023} of the interaction of a charge and a
magnet, an interaction made famous by the claims of Aharonov and
Bohm.\cite{AB1959} \ \ The following question arises. \ In an inertial frame
in which a electrically-neutral magnet is moving, are there or are there not
electric fields outside the magnet in the direction parallel to the velocity?

This puzzling situation for teachers of electromagnetism is related to the use
of continuous charge and current distributions rather than point charges.
\ Much of classical electrodynamics is taught in the historical sequence
involving continuous charge and current distributions. \ Indeed, it has been
suggested\cite{Wald} that \textit{classical} electrodynamics should deal
\textit{only} with continuous sources. \ However, if one deals with
electrodynamics from a \textit{relativistic} perspective, then suddenly point
spacetime events and point charges become important. \ Now most physicists
seem convinced that any limit may be taken and interchanged without error with
any other limit. \ Indeed, most of the time, such nonchalant interchange of
limits does not lead to errors. \ In this article, we point out an error in
connection with the interchange of Lorentz transformation and the limit of a
continuous current\ in a current loop. \ By extension, the same error occurs
in the classical electromagnetic analysis of the interaction of a
continuous-current magnet and a passing charge in different inertial frames.

\subsection{Tensor Transformations for Fields, \textit{Not} for Averages}

The explanation of the apparent relativity paradox mentioned in the first
paragraph above involves the following crucial understanding. \ The
electromagnetic field tensor assigned to a spacetime point is a mathematical
representation of a \textit{physical object} at that point, and so undergoes
tensor transformations\cite{G559}\cite{J552} between inertial frames. \ On the
other hand, \textit{averages at a single time} of electromagnetic quantities
do \textit{not} represent \textit{physical objects,} and need not undergo
tensor transformations between inertial frames. Indeed, averages at a single
time over physical quantities may vary between \textit{relativistic} inertial
frames. \ This variation in averages at a single time between
\textit{relativistic} inertial frames arises because Lorentz transformations
applied to the mathematical representations of physical quantities at
different spacetime points involve the \textit{spatial coordinates in the time
transformation}.

It should be emphasized that this \textit{variation in averages at a single
time} over physical quantities between \textit{relativistic} inertial frame is
something which does \textit{not} occur in \textit{nonrelativistic} physics,
and so seems surprising to many physicists. \ In nonrelativistic physics, time
does \textit{not} vary between inertial frames. \ Therefore a sum over the
values of some quantity at a single time in one inertial frame will correspond
to the sum over the values at a single time of the transformed quantity in a
new inertial frame. \ Like so many other \textit{relativity} paradoxes, the
relativistic mixing of space and time coordinates on Lorentz transformation
produces unfamiliar and sometimes surprising results. \ 

Furthermore, an understanding of this apparent relativity paradox suggests
again the importance of using point charges when discussing relativity and
electrodynamics. \ The limit of an electrically neutral \textit{continuous}
current loop of zero-spatial extent (an \textquotedblleft
ideal\textquotedblright\ magnetic moment) produces a magnetic field only
perpendicular to the plane of the current loop and so precludes an electric
field parallel to the direction of relative velocity between inertial frames.
\ In contrast, a current loop based upon point charges leads to a
\textit{time-varying} electric field component parallel to the relative
velocity between inertial frames. \ The average values of the time-varying
electric fields are different in different inertial frames. \ Toroids (or long
solenoids) can be regarded as stacks of current loops. \ A stack of
\textquotedblleft ideal\textquotedblright\ magnetic moments\ leads to
different electric fields from the time-varying electric fields outside a
stack of current loops based on point charges. \ The contrast in the results
of these two different models leads to different understandings of natural
phenomena. \ Indeed, the point-charge model allows a classical electromagnetic
understanding\cite{B2023} of the Aharonov-Bohm situation whereas the
\textquotedblleft ideal\textquotedblright-magnetic-moment\ model does not. \ 

\subsection{Outline of the Article}

Our analysis deals almost exclusively with the electromagnetic quantities
associated with a neutral current loop formed by a charge $q$ moving with
constant angular velocity $\omega$ in a circle of radius $R$ with an opposite
charge $-q$ at rest at the center of the loop. \ Our aim is to point out that
a loop with a \textit{continuous} current leads to a different and restricted
electric field outside the current loop compared to the \textit{point-charge}
model of the current loop. \ First we treat the charge densities, potential
functions, and electromagnetic fields and their averages for this point-charge
magnetic moment model in the inertial frame $S$ in which the circular loop is
at rest. \ Secondly, we consider these same quantities and their averages when
seen in a second $S^{\prime}$ inertial frame in which the circular loop is
moving with uniform velocity $\mathbf{V=-}\widehat{x}V$ in the $x$-direction
parallel to the plane of the loop. For simplicity in the analysis, we use the
Darwin Lagrangian approximation\cite{Darwin}\cite{PandA} and the associated
potentials and fields. Also, we assume that the orbital circle has a small
radius compared to the distance to the field point. \ The analysis suggests
the behavior of a current loop with many charges, and also the behavior of the
electromagnetic potentials and fields outside magnetic toroids and long
solenoids. \ In addition, we show that a relativity problem in Griffiths' text
involving an \textquotedblleft ideal magnetic dipole\textquotedblright\ gives
a different result from the analysis using point charges. \ We trace the
discrepancy to the use of tensor transformations for \textit{averages} of
electromagnetic quantities rather than using tensor transformations for the
actual \textit{time-varying} fields at spacetime points. \ Thus sometimes the
use of continuous charge and current densities leads to results different from
those obtained using point charges. \ Finally, we note that the relativistic
effect is unusual in involving \textit{first} order in $V/c,$ (where $c$ is
the speed of light in vacuum) like both the Fizeau experiment,\cite{Fizeau}
and the interaction of a magnet and a passing charge. \ We point out that the
classical electromagnetic analysis suggests a classical lag basis\cite{B2023}
for the Aharonov-Bohm situation when a charged particle passes through a
magnetic toroid. \ Gaussian units are used throughout the calculations. \ 

\section{Circular Orbit for a Point Charge Moving with Constant Speed}

\subsection{Exact Expressions for the Trajectory of a Point Charge}

We consider a point-charge model for a magnetic moment consisting of a point
charge $+q$ moving with constant angular velocity $\omega$ in a circle of
radius $R$ in the $xy$-plane, centered on the origin. \ A negative charge $-q$
is at rest at the origin, the center of the circle. \ \ Then we have for the
displacement of the moving charge $q$%
\begin{equation}
\mathbf{r}_{q}\left(  t\right)  =\widehat{x}R\cos\left(  \omega t+\phi
_{q}\right)  +\widehat{y}R\sin\left(  \omega t+\phi_{q}\right)  . \label{rqt}%
\end{equation}
The velocity of the charge $q$ follows as%
\begin{equation}
\mathbf{u}_{q}\left(  t\right)  =-\widehat{x}R\omega\sin\left(  \omega
t+\phi_{q}\right)  +\widehat{y}R\omega\cos\left(  \omega t+\phi_{q}\right)  ,
\label{uqt}%
\end{equation}
and the acceleration as
\begin{equation}
\mathbf{a}_{q}(t)=-\omega^{2}\mathbf{r}_{q}\left(  t\right)  =-\widehat{x}%
\omega^{2}R\cos\left(  \omega t+\phi_{q}\right)  -\widehat{y}\omega^{2}%
R\sin\left(  \omega t+\phi_{q}\right)  .
\end{equation}
The Li\'{e}nard-Wiechert expressions for the retarded potentials and fields
are given in standard textbooks of classical electrodynamics\cite{GriffithsE}%
\cite{J657} \ All of the field quantities must be evaluated at the
\textit{retarded} time $t_{q-ret}$ depending on the time, the source point,
and the field point. \ If the charge is moving and the field point is not one
of high symmetry, the calculation of the retarded time, and hence the
evaluation of the electromagnetic field, may require numerical calculation.

\subsection{Darwin-Lagrangian Approximations for a Point Charge}

In this article, our interest is simply understanding the ideas associated
with the apparent paradox noted in the opening paragraph and the validity of
going to a continuous charge or current density. \ Thus we will restrict our
discussion to small particle velocities $u_{q}=\omega R<<c$ \ for the moving
point charge $+q$ in the $S$ inertial frame, and to small relative velocities
$V<<c$ between two inertial frames $S$ and the $S^{\prime}$. \ The relative
velocity between the inertial frames is taken in the $x$-direction
$\mathbf{V=\,}\widehat{x}V$. \ We will retain terms only through order
$Vu_{q}/c^{2}$ in any expression for electromagnetic quantities. \ In this
low-speed approximation and for field points which are not too distant from
the source point, we may use the Darwin Lagrangian\cite{Darwin} and the
associated electromagnetic potentials and fields. \ The Darwin approximation
is vastly easier to work with because there are no retarded times involved.
\ The electromagnetic potentials following from the Darwin Lagrangian
are\cite{Darwin}\cite{PandA}%
\begin{equation}
\Phi(\mathbf{r},t)=\sum\nolimits_{q}\frac{q}{\left\vert \mathbf{r-r}%
_{q}(t)\right\vert }%
\end{equation}
and%
\begin{equation}
\mathbf{A}\left(  \mathbf{r,}t\right)  =\sum\nolimits_{q}\frac{q}{2c}\left[
\frac{\mathbf{u}_{q}\left(  t\right)  }{\left\vert \mathbf{r-r}_{q}%
(t)\right\vert }+\frac{\left[  \mathbf{u}_{q}\left(  t\right)  \cdot\left(
\mathbf{r-r}_{q}(t)\right)  \right]  \left(  \mathbf{r-r}_{q}(t)\right)
}{\left\vert \mathbf{r-r}_{q}(t)\right\vert ^{3}}\right]  ,
\end{equation}
and the electromagnetic fields are%

\begin{align}
\mathbf{E}\left(  \mathbf{r},t\right)   &  =\sum\nolimits_{q}q\frac{\left(
\mathbf{r-r}_{q}(t)\right)  }{\left\vert \mathbf{r-r}_{q}(t)\right\vert ^{3}%
}\left[  1+\frac{1}{2}\left(  \frac{u_{q}\left(  t\right)  }{c}\right)
^{2}-\frac{3}{2}\left(  \frac{\left(  \mathbf{r-r}_{q}(t)\right)
\cdot\mathbf{u}_{q}\left(  t\right)  }{\left\vert \mathbf{r-r}_{q}%
(t)\right\vert c}\right)  ^{2}\right] \nonumber\\
&  -\frac{q}{2c^{2}\left\vert \mathbf{r-r}_{q}(t)\right\vert }\left[
\mathbf{a}_{q}\left(  t\right)  \mathbf{+}\frac{\left[  \mathbf{a}_{q}\left(
t\right)  \mathbf{\cdot}\left(  \mathbf{r-r}_{q}(t)\right)  \right]  \left(
\mathbf{r-r}_{q}(t)\right)  }{\left\vert \mathbf{r-r}_{q}(t)\right\vert ^{2}%
}\right]  \label{EqPA}%
\end{align}
and%
\begin{equation}
\mathbf{B}\left(  \mathbf{r,t}\right)  =\sum\nolimits_{q}q\frac{\mathbf{u}%
_{q}\left(  t\right)  \mathbf{\times}\left(  \mathbf{r-r}_{q}(t)\right)
}{c\left\vert \mathbf{r-r}_{q}(t)\right\vert ^{3}}. \label{BqPA}%
\end{equation}
If for our two-charge neutral current loop we choose the field point $\left(
\mathbf{r,}t\right)  $ along the $z$-axis perpendicular to the plane of the
loop, then the retarded-time correction is a fixed quantity depending on $R$
and $z$, $t_{ret}=t-\sqrt{R^{2}+z^{2}}/c$, and it is easy to show that the
result given by the Darwin approximation agrees with the exact result through
order $\beta_{q}^{2}=u_{q}^{2}/c^{2},$ as indeed it should.

\subsubsection{First-Order in $u_{q}/c$ and $V/c$}

We are interested in an effect which is first order in the particle speed
$u_{x}/c$ and first order in the relative velocity $V/c$. \ In this
first-order approximation, we may drop the terms in the expression for the
electric field in Eq. (\ref{EqPA}) which are already of order $u_{q}/c^{2}$ or
$\omega^{2}R/c^{2}$ without involving any term in $V/c$. \ Thus we may take
the electric field as simply
\begin{equation}
\mathbf{E}\left(  \mathbf{r},t\right)  =\sum\nolimits_{q}q\frac{\left(
\mathbf{r-r}_{q}(t)\right)  }{\left\vert \mathbf{r-r}_{q}(t)\right\vert ^{3}}.
\label{ErtC}%
\end{equation}

\subsubsection{First-Order in $R/r$}

\ Finally, we assume that the radius $R$ of the current loop is small compared
to the distance to a field point $\mathbf{r=\,}\widehat{x}X+\widehat{y}Y$ in
the $xy$-plane, $R<<\sqrt{X^{2}+Y^{2},}$ so that we can use the familiar
approximations involving the electric dipole moment $\mathbf{p=}%
q\mathbf{r}_{q}$ and the magnetic dipole moment $\mathbf{m=}q\mathbf{r}%
_{q}\times\mathbf{u}_{q}/\left(  2c\right)  $. \ Then we find the potentials%
\begin{equation}
\Phi\left(  \mathbf{r},t\right)  =q\frac{\mathbf{r}_{q}\left(  t\right)
\cdot\left(  \mathbf{r-r}_{\mathbf{p}}\right)  }{\left\vert \mathbf{r-r}%
_{\mathbf{p}}\right\vert ^{3}} \label{Fdip}%
\end{equation}
and%
\begin{equation}
\mathbf{A}\left(  \mathbf{r},t\right)  =q\left[  \frac{\mathbf{r}_{q}\left(
t\right)  \times\mathbf{u}_{q}\left(  t\right)  }{2c}\right]  \frac
{\times\left(  \mathbf{r-r}_{\mathbf{p}}\right)  }{\left\vert \mathbf{r-r}%
_{\mathbf{p}}\right\vert ^{3}} \label{Adip}%
\end{equation}
where $\mathbf{r}_{\mathbf{p}}$ is the location of the center of the current
loop. \ In this approximation, the fields are given by dipole approximations
as
\begin{equation}
\mathbf{E}\left(  \mathbf{r,}t\right)  =q\left(  \frac{3\left[  \mathbf{r}%
_{q}\left(  t\right)  \cdot\left(  \mathbf{r-r}_{\mathbf{p}}\right)  \right]
\left(  \mathbf{r-r}_{\mathbf{p}}\right)  }{\left\vert \mathbf{r-r}%
_{\mathbf{p}}\right\vert ^{5}}-\frac{\mathbf{r}_{q}\left(  t\right)
}{\left\vert \mathbf{r-r}_{\mathbf{p}}\right\vert ^{3}}\right)  \label{Edip}%
\end{equation}
and
\begin{equation}
\mathbf{B}\left(  \mathbf{r},t\right)  =q\left\{  3\left[  \frac{\left(
\mathbf{r}_{q}\times\mathbf{u}_{q}\right)  }{2c}\right]  \cdot\frac{\left(
\mathbf{r-r}_{\mathbf{p}}\right)  }{\left\vert \mathbf{r-r}_{\mathbf{p}%
}\right\vert ^{5}}-\left[  \frac{\left(  \mathbf{r}_{q}\times\mathbf{u}%
_{q}\right)  }{2c}\right]  \frac{1}{\left\vert \mathbf{r-r}_{\mathbf{p}%
}\right\vert ^{3}}\right\}  . \label{Bdip}%
\end{equation}

\subsection{Time-Varying Electric Quantities in the $S$ Inertial Frame}

The position $\mathbf{r}_{q}\left(  t\right)  $ and the velocity
$\mathbf{u}_{q}\left(  t\right)  $ of the moving point charge are varying in
time and lead to time-variations for the electric potential in Eq.
(\ref{Fdip}) and the electric field in Eq. (\ref{Edip}) with the center of the
current loop at the origin, $\mathbf{r}_{\mathbf{p}}=0$. \ On the other hand
the velocity $\mathbf{u}_{q}\left(  t\right)  $ of the charge $q$ is always
perpendicular to the displacement $\mathbf{r}_{q}\left(  t\right)  $ so that
$\mathbf{r}_{q}\left(  t\right)  \times\mathbf{u}_{q}\left(  t\right)
=\widehat{z}Ru_{q}=\widehat{z}\omega R^{2}$ is constant in time. \ Accordingly
the vector potential and the magnetic field are both constant in time as
\begin{equation}
\mathbf{A}\left(  \mathbf{r},t\right)  =\mathbf{A}\left(  \mathbf{r}\right)
=\frac{\mathbf{m}\times\mathbf{r}}{r^{3}}%
\end{equation}
and
\begin{equation}
\mathbf{B}\left(  \mathbf{r},t\right)  =\mathbf{B}\left(  \mathbf{r}\right)
=\left\{  \frac{3\left(  \mathbf{m\cdot r}_{q}\right)  \mathbf{r}_{q}}{r^{5}%
}-\frac{\mathbf{m}}{r^{3}}\right\}
\end{equation}
where the magnetic moment $\mathbf{m}$ is constant in time%
\begin{equation}
\mathbf{m}\left(  t\right)  =q\left[  \frac{\left(  \mathbf{r}_{q}%
\times\mathbf{u}_{q}\right)  }{2c}\right]  =\widehat{z}q\frac{Ru_{q}}%
{2c}=\widehat{z}q\frac{\omega R^{2}}{2c}=\widehat{z}m.
\end{equation}

\subsection{Averages and the \textquotedblleft Ideal\textquotedblright%
\ Magnetic Dipole Moment}

\subsubsection{Averages at time $t$ for the Point-Charge Model}

Although the electric potential and electric field of our
\textit{point-charge} magnetic moment model are \textit{varying} in time,
these electric quantities vanish when averaged over the phase $\phi_{q}$ at a
single time $t$ in the $S$ inertial frame
\begin{equation}
\left\langle \mathbf{r}_{q}\left(  t\right)  \right\rangle _{\phi_{q}%
}=0,\text{ \ \ }\left\langle \Phi\left(  \mathbf{r},t\right)  \right\rangle
_{\phi_{q}}=0,\text{ \ \ and \ \ }\left\langle \mathbf{E}\left(
\mathbf{r},t\right)  \right\rangle _{\phi_{q}}=0.
\end{equation}

\subsubsection{The \textquotedblleft Ideal\textquotedblright\ Magnetic Dipole
Moment}

This vanishing of the averages of the time-varying electric quantities arising
in the point-charge model for a magnetic moment has suggested the idea of an
\textquotedblleft ideal\textquotedblright\ magnetic dipole moment. \ The
average over the phase angle $\phi_{q}$ has the same effect as considering
many non-interacting point charges in a current loop and then taking the
continuous current limit of many point charges subdivided repeatedly while
maintaining the same magnetic moment. \ Indeed, the time-independent magnetic
dipole moment $\mathbf{m}$ is the only non-vanishing quantity for our
small-radius magnetic moment. \ If we imagine the radius $R$ of the current
loop as negligible in size, then we have an \textquotedblleft
ideal\textquotedblright\ point magnetic dipole of magnetic moment $\mathbf{m}$
and magnetization $\mathbf{M(r)=m}\delta^{3}\left(  \mathbf{r}\right)  $.
\ The associated current density is $\mathbf{J}_{\mathbf{m}}\left(
\mathbf{r}\right)  =c\nabla\times\mathbf{M}\left(  \mathbf{r}\right)
=c\mathbf{m\times}\nabla\delta^{3}\left(  \mathbf{r}\right)  ,$ while the
charge density vanishes $\rho_{\mathbf{m}}\left(  \mathbf{r}\right)  =0,$
which is consistent with the charge continuity equation $\partial
\rho_{\mathbf{m}}/\partial t=-\nabla\cdot\mathbf{J}_{\mathbf{m}}$. \ In the
$xy$-plane containing the \textquotedblleft ideal\textquotedblright\ magnetic
moment, the only magnetic field will be in the $\widehat{\mathbf{z}}%
$-direction. \ 

\subsubsection{\textquotedblleft Ideal\textquotedblright\ Magnetic Toroids and
Solenoids}

We can imagine forming magnetic toroids and magnetic solenoids as stacks of
\textquotedblleft ideal\textquotedblright\ magnetic dipoles. \ From this
perspective, there are no electric or magnetic field outside a \textit{toroid}
formed from such \textquotedblleft ideal\textquotedblright\ magnetic moments.
\ Indeed, one can consider a field point along the axis of a \textit{magnetic
toroid} formed from such \textquotedblleft ideal\textquotedblright\ magnetic
and can conclude by symmetry alone, that the magnetic field must vanish.
\ What we wish to point out is that this \textquotedblleft
ideal\textquotedblright\ magnetic dipole moment involves different results
compared to the point-charge model for the electric field in an inertial frame
$S^{\prime}$ in which the magnetic moment is moving. \ 

\section{Analysis of the Point-Charge Magnetic Moment in the $S^{\prime}$
Inertial Frame}

\subsection{Trajectory Equations for the Moving Charge $q$ in the Inertial
Frame $S^{\prime}$}

We will be considering relativistic transformations of the electromagnetic
quantities between the inertial frames labeled by $S$ and $S^{\prime}$. \ We
will take the $S^{\prime}$ inertial frame as moving with constant velocity
$\mathbf{V=}\widehat{x}V$ relative to the $S$ inertial frame. \ The
displacement of the charge $q$ in the $S^{\prime}$ inertial frame is obtained
by Lorentz transformations $x^{\prime}=\Gamma\left(  x-Vt\right)  $ and
$y^{\prime}=y,$ $z^{\prime}=z,~$and $t^{\prime}=\Gamma(t-Vx/c^{2})$\ where
$\Gamma=\left[  1-\left(  V/c\right)  ^{2}\right]  ^{-1/2}$ giving \
\begin{equation}
\mathbf{r}_{q}^{\prime}\left(  t^{\prime}\right)  =\widehat{x}\left\{  \Gamma
R\cos\left[  \omega t+\phi_{q}\right]  -Vt\right\}  +\widehat{y}R\sin\left[
\omega t+\phi_{q}\right]  .
\end{equation}
We note that $x_{q}=\Gamma\left(  x_{q}^{\prime}+Vt^{\prime}\right)  $ and use
the Lorentz transformation for the time $t=\Gamma(t^{\prime}+Vx^{\prime}%
/c^{2})$ to eliminate the unprimed time $t$ in favor of primed quantities,
\begin{equation}
x_{q}^{\prime}=-Vt^{\prime}+\frac{1}{\Gamma}R\cos\left[  \omega\Gamma
(t^{\prime}+Vx_{q}^{\prime}/c^{2})+\phi_{q}\right]  \label{xpp}%
\end{equation}
and
\begin{equation}
y_{q}^{\prime}=R\sin\left[  \omega\Gamma(t^{\prime}+Vx_{q}^{\prime}%
/c^{2})+\phi_{q}\right]  . \label{ypp}%
\end{equation}
These equations give the exact trajectory for the charge in the $S^{\prime}$
inertial frame. \ Thus, in the $S^{\prime}$ inertial frame, we have
\textit{implicit} functions for the coordinates $x_{q}^{\prime}$ and
$y_{q}^{\prime}$ in terms of the time $t^{\prime}$. \ However, there seems to
be no closed-form explicit solution for these equations. \ Thus, the situation
here is quite different from that for a charged particle moving with constant
velocity where the trajectory in a different relativistic inertial frame can
be given in closed form.\cite{J657}

\subsection{Trajectory of the Moving Charge in the Darwin-Lagrangian
Approximation}

Although there is no exact closed-form expression for the trajectory of the
charge $q$ in the $S^{\prime}$ inertial frame, we can find an approximate
expression within the approximations we have introduced. \ Thus if we keep
only terms through first order in $V/c$ and first order in $u_{q}/c$, and
through first order in $r_{q}/r$, then we find for the trajectory of the
moving charge,%
\begin{equation}
x_{q}^{\prime}\left(  t^{\prime}\right)  =\Gamma\left(  x_{q}-Vt\right)
\approx x_{q}\left(  t\right)  -Vt=x_{q}\left(  t\right)  -Vt^{\prime},\text{
\ }y_{q}^{\prime}\left(  t^{\prime}\right)  =y_{q}\left(  t\right)  ,\text{
\ }z_{q}^{\prime}=z_{q}=0 \label{xpqtpy}%
\end{equation}
and
\begin{equation}
t_{q}^{\prime}=\Gamma\left(  t-Vx_{q}\left(  t\right)  /c^{2}\right)  \approx
t-\left(  V/c\right)  x_{q}\left(  t\right)  /c=t-\left(  V/c\right)
x_{q}\left(  t^{\prime}\right)  /c, \label{tpqxq}%
\end{equation}
and where the displacement $x_{q}\left(  t\right)  $ in the unprimed $S$
inertial frame is given in Eq. (\ref{rqt}). \ We notice that the approximate
transformation involving first-order terms for the relative speed $V$ gives a
transformation for the displacement $x_{q}^{\prime}$ in Eq. (\ref{xpqtpy})
which is the same as is involved for nonrelativistic physics. \ It is the time
transformation which is completely different. \ The time transformation is
first-order in the relative velocity $V$ between the $S$ and $S^{\prime}$
inertial frames but is of order $1/c^{2}$. \ Since we are interested in
averages over the phase $\phi_{q}$ (or equivalently averaging over many
charged particles) at a single time $t$ in the $S$ inertial frame or at a
single time $t^{\prime}$ in the $S^{\prime}$ inertial frame, the time
transformation may lead to quite different results for averages in different
inertial frames.

\subsection{Electric Dipole Moment for the Charges in $S^{\prime}$}

In the $S^{\prime}$ inertial frame the charge $q$ has the location
$x_{q}^{\prime}\left(  t^{\prime}\right)  =x_{q}\left(  t\right)
-Vt=x_{q}\left(  t\right)  -Vt^{\prime},y_{q}^{\prime}\left(  t^{\prime
}\right)  =y_{q}\left(  t\right)  $ at time $t^{\prime}=t-\left(  V/c\right)
x_{q}\left(  t\right)  /c=t-\left(  V/c\right)  x_{q}\left(  t^{\prime
}\right)  /c$. \ Then $x_{q}^{\prime}(t^{\prime})+Vt^{\prime}=x_{q}(t)$ or
keeping only first-order terms in $V/c$, we have
\begin{align}
x_{q}^{\prime}(t^{\prime})+Vt^{\prime}  &  =x_{q}\left[  t^{\prime}%
+\frac{\left(  V\right)  }{c}\frac{x_{q}(t)}{c}\right] \nonumber\\
&  \approx R\cos\left\{  \omega_{q}\left[  t^{\prime}+\frac{\left(  V\right)
}{c}\frac{x_{q}(t)}{c}\right]  +\phi_{q}\right\} \nonumber\\
&  =R\cos\left(  \omega_{q}t^{\prime}+\phi_{q}\right)  \cos\left[  \omega
_{q}\frac{\left(  V\right)  }{c}\frac{x_{q}(t)}{c}\right] \nonumber\\
&  -R\sin\left(  \omega_{q}t^{\prime}+\phi_{q}\right)  \sin\left[  \omega
_{q}\frac{\left(  V\right)  }{c}\frac{x_{q}(t)}{c}\right] \nonumber\\
&  =R\cos\left(  \omega_{q}t^{\prime}+\phi_{q}\right)  -\frac{\left(
V\right)  }{c}\frac{\omega_{q}R}{c}R\cos\left(  \omega_{q}t^{\prime}\right)
\sin\left(  \omega_{q}t^{\prime}+\phi_{q}\right)  \label{xpqt}%
\end{align}
and
\begin{align}
y_{q}^{\prime}(t^{\prime})  &  =y_{q}(t)=y\left[  t^{\prime}+\frac{\left(
V\right)  }{c}\frac{x_{q}(t)}{c}\right] \nonumber\\
&  =R\sin\left\{  \omega_{q}\left[  t^{\prime}+\frac{\left(  V\right)  }%
{c}\frac{x_{q}(t)}{c}\right]  +\phi_{q}\right\} \nonumber\\
&  =R\sin\left(  \omega_{q}t^{\prime}+\phi_{q}\right)  \cos\left[  \omega
_{q}\frac{\left(  V\right)  }{c}\frac{x_{q}(t)}{c}\right] \nonumber\\
&  +R\cos\left(  \omega_{q}t^{\prime}+\phi_{q}\right)  \sin\left[  \omega
_{q}\frac{\left(  V\right)  }{c}\frac{x_{q}(t)}{c}\right] \nonumber\\
&  =R\sin\left(  \omega_{q}t^{\prime}+\phi_{q}\right)  +\frac{\left(
V\right)  }{c}\frac{\omega_{q}R}{c}R\cos^{2}\left(  \omega_{q}t^{\prime}%
+\phi_{q}\right)  , \label{ypqt}%
\end{align}
where we have noted that $\sin\left(  a+bx/c^{2}\right)  \approx\sin a+\left(
bx/c^{2}\right)  \cos a$ and $\cos\left(  a+bx/c^{2}\right)  \approx\cos
a-\left(  bx/c^{2}\right)  \sin a$ through order $1/c^{2}$. \ Our results are
first order in $V/c$ and first order in $u_{q}/c=\omega R/c$. \ 

If we calculate the $\phi_{q}$-average electric dipole moment at time
$t^{\prime}$, we find $\mathbf{p}^{\prime}\left(  t^{\prime}\right)
=q\mathbf{r}_{q}^{\prime}\left(  t^{\prime}\right)  =\widehat{x}qx_{q}%
^{\prime}\left(  t^{\prime}\right)  +\widehat{y}qy_{q}^{\prime}\left(
t^{\prime}\right)  ~\ $gives an average at time $t^{\prime}$%
\begin{equation}
\left\langle \mathbf{p}^{\prime}\left(  t^{\prime}\right)  \right\rangle
=\left\langle q\mathbf{r}_{q}^{\prime}\left(  t^{\prime}\right)  \right\rangle
=\widehat{x}q\left\langle x_{q}^{\prime}\left(  t^{\prime}\right)
\right\rangle +\widehat{y}q\left\langle y_{q}^{\prime}\left(  t^{\prime
}\right)  \right\rangle =\widehat{y}\frac{V\omega_{q}R^{2}}{2c^{2}},
\label{Appt'}%
\end{equation}
which does not vanish in $S^{\prime}$. \ The electric dipole moment of the
$\pm q$ current loop has vanishing average value in $S$ but non-zero average
in $S^{\prime}$. \ Thus, indeed averages over extended electromagnetic systems
at a fixed time $t$ or $t^{\prime}$ can vary with the choice of inertial frame.

\subsection{Scalar Potential and Electric Field $E_{x}$ in the $S^{\prime}$
Inertial Frame}

All the approximations which held in the $S$ inertial frame are also valid in
the $S^{\prime}$ inertial frame. \ Thus, in the $S^{\prime}$ inertial frame,
the scalar potential due to the charge $\pm q$ is given by
\begin{equation}
\Phi^{\prime}\left(  \mathbf{r}^{\prime},t^{\prime}\right)  =\frac
{\mathbf{p}^{\prime}\left(  t^{\prime}\right)  \cdot\left(  \mathbf{r}%
^{\prime}-\mathbf{r}_{\mathbf{p}}^{\prime}\right)  }{\left\vert \mathbf{r}%
^{\prime}-\mathbf{r}_{\mathbf{p}}^{\prime}\right\vert ^{3}}=\frac
{q\mathbf{r}^{\prime}\left(  t^{\prime}\right)  \cdot\left(  \mathbf{r}%
^{\prime}-\mathbf{r}_{\mathbf{p}}^{\prime}\right)  }{\left\vert \mathbf{r}%
^{\prime}-\mathbf{r}_{\mathbf{p}}^{\prime}\right\vert ^{3}}%
\end{equation}
with the center of the dipole located at $\mathbf{r}_{\mathbf{p}}^{\prime
}=-\mathbf{V}T^{\prime}$. \ If we take the field point in the $xy$-plane at
$X^{\prime},Y^{\prime},0,T^{\prime},$ this becomes%
\begin{equation}
\Phi^{\prime}\left(  X^{\prime},Y^{\prime},0,T^{\prime}\right)  =q\frac
{\left(  X^{\prime}+VT^{\prime}\right)  x_{a}^{\prime}\left(  T^{\prime
}\right)  +Y^{\prime}y_{q}^{\prime}\left(  T^{\prime}\right)  }{\left[
\left(  X^{\prime}+VT^{\prime}\right)  +Y^{^{\prime2}}\right]  ^{3/2}},
\label{FpXpq}%
\end{equation}
where $x_{q}^{\prime}$ and $y_{q}^{\prime}$ are given in Eqs. (\ref{xpqt}) and
(\ref{ypqt}). \ The $\phi_{q}$-average value at time $T^{\prime}$ is
\begin{align}
\left\langle \Phi^{\prime}\left(  X^{\prime},Y^{\prime},0,T^{\prime}\right)
\right\rangle _{\phi_{q}}  &  =q\frac{\left(  X^{\prime}+VT^{\prime}\right)
\left\langle x_{a}^{\prime}\left(  T^{\prime}\right)  \right\rangle
+Y^{\prime}\left\langle y_{q}^{\prime}\left(  T^{\prime}\right)  \right\rangle
}{\left[  \left(  X^{\prime}+VT^{\prime}\right)  ^{2}+Y^{^{\prime2}}\right]
^{3/2}}\nonumber\\
&  =q\frac{V\omega_{q}R^{2}}{2c^{2}}\frac{Y^{\prime}}{\left[  \left(
X^{\prime}+VT^{\prime}\right)  ^{2}+Y^{^{\prime2}}\right]  ^{3/2}}
\label{AvFpXp}%
\end{align}
from Eq. (\ref{Appt'}).

The electric field in $S^{\prime}$ is given by
\begin{equation}
\mathbf{E}^{\prime}\left(  \mathbf{r}^{\prime}\mathbf{,}t^{\prime}\right)
=q\left(  \frac{3\left[  \mathbf{r}_{q}^{\prime}\left(  t^{\prime}\right)
\cdot\left(  \mathbf{r}^{\prime}-\mathbf{r}_{\mathbf{p}}^{\prime}\right)
\right]  \left(  \mathbf{r}^{\prime}-\mathbf{r}_{\mathbf{p}}^{\prime}\right)
}{\left\vert \mathbf{r}^{\prime}-\mathbf{r}_{\mathbf{p}}^{\prime}\right\vert
^{5}}-\frac{\mathbf{r}_{q}^{\prime}\left(  t^{\prime}\right)  }{\left\vert
\mathbf{r}^{\prime}-\mathbf{r}_{\mathbf{p}}^{\prime}\right\vert ^{3}}\right)
\end{equation}
or%
\begin{align}
\mathbf{E}^{\prime}\left(  X^{\prime},Y^{\prime},0,T^{\prime}\right)   &
=q\left(  \frac{3\left[  x_{q}^{\prime}\left(  t^{\prime}\right)  \left(
X^{\prime}+VT^{\prime}\right)  +y_{q}^{\prime}\left(  t^{\prime}\right)
Y^{\prime}\right]  \left[  \widehat{x}\left(  X^{\prime}+VT^{\prime}\right)
+\widehat{y}Y^{\prime}\right]  }{\left[  \left(  X^{\prime}+VT^{\prime
}\right)  ^{2}+Y^{^{\prime2}}\right]  ^{5/2}}\right. \nonumber\\
&  \left.  -\frac{\widehat{x}x_{q}^{\prime}\left(  t^{\prime}\right)
+\widehat{y}y_{q}^{\prime}\left(  t^{\prime}\right)  }{\left[  \left(
X^{\prime}-VT^{\prime}\right)  ^{2}+Y^{^{\prime2}}\right]  ^{3/2}}\right)
\end{align}
where the moving dipole has its center at $\mathbf{r}_{\mathbf{p}}^{\prime
}\left(  t^{\prime}\right)  =-\mathbf{V}t^{\prime}$. \ The $\phi_{q}$-average
at time $T^{\prime}$ is%
\begin{align}
\left\langle \mathbf{E}^{\prime}\left(  X^{\prime},Y^{\prime},0,T^{\prime
}\right)  \right\rangle _{\phi_{q}}  &  =q\left(  \frac{3\left[  \left\langle
x_{q}^{\prime}\left(  t^{\prime}\right)  \right\rangle \left(  X^{\prime
}+VT^{\prime}\right)  +\left\langle y_{q}^{\prime}\left(  t^{\prime}\right)
\right\rangle Y^{\prime}\right]  \left[  \widehat{x}\left(  X^{\prime
}+VT^{\prime}\right)  +\widehat{y}Y^{\prime}\right]  }{\left[  \left(
X^{\prime}+VT^{\prime}\right)  ^{2}+Y^{^{\prime2}}\right]  ^{5/2}}\right.
\nonumber\\
&  \left.  -\frac{\widehat{x}\left\langle x_{q}^{\prime}\left(  t^{\prime
}\right)  \right\rangle +\widehat{y}\left\langle y_{q}^{\prime}\left(
t^{\prime}\right)  \right\rangle }{\left[  \left(  X^{\prime}+VT^{\prime
}\right)  ^{2}+Y^{^{\prime2}}\right]  ^{3/2}}\right) \nonumber\\
&  =\left[  \frac{V\omega_{q}R^{2}}{2c^{2}}\right]  \left(  3\frac{Y^{\prime
}\left[  \widehat{x}\left(  X^{\prime}+VT^{\prime}\right)  ^{\prime
}+\widehat{y}Y^{\prime}\right]  }{\left[  \left(  X^{\prime}+VT^{\prime
}\right)  ^{2}+Y^{^{\prime2}}\right]  ^{5/2}}-\frac{\widehat{y}}{\left[
\left(  X^{\prime}+VT^{\prime}\right)  ^{2}+Y^{^{\prime2}}\right]  ^{3/2}%
}\right) \nonumber\\
&  =\left[  \frac{V\omega_{q}R^{2}}{2c^{2}}\right]  \left(  3\frac{Y^{\prime
}\left[  \widehat{x}X^{\prime}+\widehat{y}Y^{\prime}\right]  }{\left[  \left(
X^{\prime}\right)  ^{2}+Y^{^{\prime2}}\right]  ^{5/2}}-\frac{\widehat{y}%
}{\left[  \left(  X^{\prime}\right)  ^{2}+Y^{^{\prime2}}\right]  ^{3/2}%
}\right)  \label{EpXpYpTp}%
\end{align}
again from Eq. (\ref{Appt'}). \ 

\subsection{The Vector Potential and Magnetic Field in $S^{\prime}$}

Since the vector potential is already first order in $1/c$, we may use
nonrelativistic transformations between $S$ and $S^{\prime}$ when dealing with
the vector potential. \ The vector potential is simply%

\begin{equation}
\mathbf{A}^{\prime}\left(  X^{\prime},Y^{\prime},0,T^{\prime}\right)
=m\frac{\mathbf{-}\widehat{x}Y^{\prime}+\widehat{y}\left(  X^{\prime
}+VT^{\prime}\right)  }{\left[  \left(  X^{\prime}+VT^{\prime}\right)
^{2}+Y^{^{\prime2}}\right]  ^{3/2}}, \label{ApXpYp0}%
\end{equation}
where the magnetic moment is unchanged,%
\begin{equation}
\mathbf{m}^{\prime}\left(  t\right)  =q\left[  \frac{\left(  \mathbf{r}%
_{q}^{\prime}\times\mathbf{u}_{q}^{\prime}\right)  }{2c}\right]
=\widehat{z}q\frac{\omega R^{2}}{2c}=\widehat{z}m.
\end{equation}
since the expression for $\mathbf{m}$ is already first order in $1/c$ and we
are dropping terms in $1/c^{3}$. \ \ The vector potential leads to a magnetic
field which for field points $\mathbf{\ r,}t$ in $S$ or $\mathbf{r}^{\prime
},t^{\prime}$ in $S^{\prime}$ is purely in the $z$-direction\ and%
\begin{equation}
\mathbf{B}\left(  X^{\prime},Y^{\prime},0,T^{\prime}\right)  =-\widehat{z}%
\frac{m}{\left[  \left(  X^{\prime}+VT^{\prime}\right)  ^{2}+Y^{^{\prime2}%
}\right]  ^{3/2}}.
\end{equation}

\subsection{Lorentz Transformation of the Average Values}

We saw above that in the $S$ inertial frame, the $\phi_{q}$-average values at
a fixed time $t$ for the point-charge model for a magnetic dipole agreed with
the values given for the \textquotedblleft ideal\textquotedblright\ magnetic
dipole moment. \ If we carry out Lorentz transformations for the
\textquotedblleft ideal\textquotedblright\ magnetic moment from the $S$ to the
$S^{\prime}~$inertial frame, we find%
\begin{equation}
\Phi_{\mathbf{m}}^{\prime}\left(  \mathbf{r}^{\prime},t^{\prime}\right)
\approx0-\frac{V}{c}A_{x}\left(  \mathbf{r,}t\right)
\end{equation}
giving%
\begin{align}
\Phi_{\mathbf{m}}^{\prime}\left(  X^{\prime},Y^{\prime},0,T^{\prime}\right)
&  \approx0-\frac{V}{c}m\frac{-Y}{\left(  X^{2}+Y^{2}\right)  ^{3/2}%
}\nonumber\\
&  =-\frac{V}{c}m\frac{-Y^{\prime}}{\left[  \left(  X^{\prime}+VT^{\prime
}\right)  ^{2}+Y^{\prime2}\right]  ^{3/2}}\nonumber\\
&  \approx\frac{V}{c}m\frac{Y^{\prime}}{\left[  X^{\prime2}+Y^{\prime
2}\right]  ^{3/2}} \label{FmXp}%
\end{align}
through first order in $V/c$. \ We notice that this result for the scalar
potential agrees with the \textit{average} value for the time-varying scalar
potential given in Eq. (\ref{AvFpXp}). \ However, the average value does not
include the additional time-varying term involving $x_{q}^{\prime}\left(
t^{\prime}\right)  $ in Eq. (\ref{FpXpq}). \ It is this additional
time-varying term which, when combined with the relativistic time dependence,
gives an average value for the electric field parallel to the velocity of the
moving current loop in the $S^{\prime}$ inertial frame. \ 

Since the $\phi_{q}$-average scalar potential vanishes in $S$, the $\phi_{q}%
$-average vector potential (within our approximations) is given by
$\mathbf{A}^{\prime}\left(  \mathbf{r}^{\prime},t^{\prime}\right)
\approx\mathbf{A}\left(  \mathbf{r,}t\right)  ,$ since the expression is
already first order in $u_{q}/c=\omega R/c$ and we are dropping any terms in
$1/c^{3}$. \ Since the only corrections to the non-relativistic expressions
are already of order $1/c^{2}$ in Eq. (\ref{tpqxq}), we have%
\begin{equation}
\mathbf{A}_{\mathbf{m}}^{\prime}\left(  X^{\prime},Y^{\prime},0,T^{\prime
}\right)  \approx m\frac{\mathbf{-}\widehat{x}Y^{\prime}+\widehat{y}\left(
X^{\prime}+VT^{\prime}\right)  }{\left[  \left(  X^{\prime}+VT^{\prime
}\right)  ^{2}+Y^{^{\prime2}}\right]  ^{3/2}}. \label{ApXp}%
\end{equation}
Working from the \textquotedblleft ideal\textquotedblright\ magnetic dipole
potentials in Eqs. (\ref{FmXp}) and (\ref{ApXp}), we find the magnetic field
from $B=\nabla\times\mathbf{A}$ giving
\begin{equation}
\mathbf{B}_{\mathbf{m}}^{\prime}\left(  X^{\prime},Y^{\prime},0,T^{\prime
}\right)  =-\widehat{z}\frac{m}{\left[  \left(  X^{\prime}+VT^{\prime}\right)
^{2}+Y^{^{\prime2}}\right]  ^{3/2}},
\end{equation}
and the electric field from $\mathbf{E=-}\nabla\Phi-(1/c)\partial
\mathbf{A/\partial}t$ giving%
\begin{align}
\mathbf{E}_{\mathbf{m}}^{\prime}\left(  X^{\prime},Y^{\prime},0,T^{\prime
}\right)   &  =-\frac{V}{c}m\left\{  \frac{-\widehat{y}}{\left[  \left(
\left(  X^{\prime}+VT^{\prime}\right)  ^{2}\right)  ^{2}+Y^{\prime2}\right]
^{3/2}}-3\frac{\widehat{x}\left(  X^{\prime}+VT^{\prime}\right)  Y^{\prime
}+\widehat{y}\left(  Y^{\prime}\right)  ^{2}}{\left[  \left(  \left(
X^{\prime}+VT^{\prime}\right)  ^{2}\right)  ^{2}+Y^{\prime2}\right]  ^{5/2}%
}\right\} \nonumber\\
&  -\frac{1}{c}\frac{\partial}{\partial T^{\prime}}\left(  m\frac
{\mathbf{-}\widehat{x}Y^{\prime}+\widehat{y}\left(  X^{\prime}+VT^{\prime
}\right)  }{\left[  \left(  X^{\prime}+VT^{\prime}\right)  ^{2}+Y^{^{\prime2}%
}\right]  ^{3/2}}\right) \nonumber\\
&  =\frac{V}{c}m\frac{\widehat{y}}{\left[  \left(  \left(  X^{\prime
}+VT^{\prime}\right)  ^{2}\right)  ^{2}+Y^{\prime2}\right]  ^{3/2}}%
+3\frac{\widehat{x}\left(  X^{\prime}+VT^{\prime}\right)  Y^{\prime
}+\widehat{y}\left(  Y^{\prime}\right)  ^{2}}{\left[  \left(  \left(
X^{\prime}+VT^{\prime}\right)  ^{2}\right)  ^{2}+Y^{\prime2}\right]  ^{5/2}%
}\nonumber\\
&  -\frac{Vm}{c}\left\{  \frac{\widehat{y}}{\left[  \left(  X^{\prime
}+VT^{\prime}\right)  ^{2}+Y^{^{\prime2}}\right]  ^{3/2}}-3\frac{\left[
\mathbf{-}\widehat{x}Y^{\prime}+\widehat{y}\left(  X^{\prime}+VT^{\prime
}\right)  \right]  \left(  X^{\prime}+VT^{\prime}\right)  }{\left[  \left(
\left(  X^{\prime}+VT^{\prime}\right)  ^{2}\right)  ^{2}+Y^{\prime2}\right]
^{5/2}}\right\} \nonumber\\
&  =\frac{V}{c}m\left\{  \frac{\widehat{y}}{\left[  \left(  \left(  X^{\prime
}\right)  ^{2}\right)  ^{2}+Y^{\prime2}\right]  ^{3/2}}+3\frac{\widehat{x}%
\left(  X^{\prime}\right)  Y^{\prime}+\widehat{y}\left(  Y^{\prime}\right)
^{2}}{\left[  \left(  \left(  X^{\prime}\right)  ^{2}\right)  ^{2}+Y^{\prime
2}\right]  ^{5/2}}\right\} \nonumber\\
&  -\frac{Vm}{c}\left\{  \frac{\widehat{y}}{\left[  \left(  X^{\prime}\right)
^{2}+Y^{^{\prime2}}\right]  ^{3/2}}-3\frac{\left[  \mathbf{-}\widehat{x}%
Y^{\prime}+\widehat{y}\left(  X^{\prime}\right)  \right]  \left(  X^{\prime
}\right)  }{\left[  \left(  X^{\prime}\right)  ^{2}+Y^{\prime2}\right]
^{5/2}}\right\} \nonumber\\
&  =\widehat{y}\frac{V}{c}m\left\{  \frac{3\left(  X^{\prime2}+Y^{\prime
2}\right)  }{\left[  \left(  X^{\prime}\right)  ^{2}+Y^{\prime2}\right]
^{5/2}}\right\}  =\widehat{y}\frac{V}{c}m\left\{  \frac{3}{\left[  \left(
X^{\prime}\right)  ^{2}+Y^{\prime2}\right]  ^{3/2}}\right\}  .
\end{align}

\subsection{The \textquotedblleft Ideal\textquotedblright\ Magnetic Moment
Model Leaves Out Terms in the Electric Potential and Electric Field}

We notice that if we use the $\phi_{q}$-average value of the point-charge
model or the \textquotedblleft ideal\textquotedblright\ magnetic moment model
in the $S$ inertial frame, there is no component of the electric field in the
direction of the relative velocity in the $S^{\prime}$ inertial frame. \ The
tensor transformation $E_{x}=E_{x}^{\prime}$ for the electromagnetic
fields\cite{G559} \cite{J552} shows this same discrepancy depending upon
whether or not one takes the average values in the $S$ inertial frame. \ If
one takes the average before making the Lorentz transformation to the
$S^{\prime}$ inertial frame, one loses the time-varying electric dipole which
gives a time-varying scalar potential and a time-varying electric field
$E_{x}$ in the $x$-direction. \ The time-varying expressions in the new
$S^{\prime}$ inertial frame then lead to new \textit{average} values in this
frame at a single time $t^{\prime}$ because of the space-coordinate dependence
of the time transformation given in Eq. (\ref{tpqxq}). \ Average values at a
fixed time lead to different averages in different inertial frames. \ This
inertial-frame dependence of the averages is strikingly illustrated by the
absence of any electric dipole moment for our point-charge system in the $S$
inertial frame and the existence of a non-zero average dipole moment for our
system in the $S^{\prime}$ inertial frame. \ The electromagnetic field tensor
at a spacetime point is a mathematical representation of a physical object and
the reprentation has tensor transformations between inertial frames.
\ Averages at a single time have no physical existence but depend upon the
choice of inertial frame in which the average is evaluated.

\section{Comments On the Analysis}

\subsection{Straight Line Current and Point-Charge Models}

When dealing with relativity and electrodynamics, all textbooks discuss a
straight line current. \ In this case, Lorentz transformations do not betray
the importance of using point-charge models for the currents. \ The order of
Lorentz transformations and the limit to continuous currents makes no
difference. \ Of course, when trying to give a physical picture of what is
involved in the sudden appearance of a non-zero charge density in a moving
inertial frame $S^{\prime}$ from a neutral wire in the electrically neutral
wire, the textbook discussion retreats from the continuous-current limit over
to the point-charge picture. \ The textbook makes the Lorentz transformation
in the point-charge picture, and then goes back to the continuous-current
limit. \ 

\subsection{Failures of the Continuous-Current Model for a Magnet Moment}

The straight-line continuous current which appears in all the textbooks
betrays no error in interchanging the continuous-current limit and Lorentz
transformation. \ However, in the case of a current loop, the interchange of
the order of continuous-current limits and Lorentz transformation is not
successful. \ Use of the continuous limit before making the Lorentz
transformation leaves out an important part of the \textit{time-varying}
electric potential and \textit{time-varying} electric field. \ The use of the
average potential or average electric field in the $S$ inertial frame leads to
the \textquotedblleft ideal\textquotedblright\ magnetic moment which has no
electric field. \ Then use of Lorentz transformation omits the electric field
parallel to the relative velocity between the relativistic inertial frames.
\ Thus, the interchange of averages and Lorentz transformations fails for this reason.

The use of the \textquotedblleft ideal\textquotedblright\ magnetic moment
model also fails for another reason. \ The \textquotedblleft
ideal\textquotedblright\ magnetic moment may start with a continuous current
$I$ and finite radius~$R$, giving a magnetic moment of magnitude $m=I\pi
R^{2}/c$, but it takes the limit to a very small (zero) radius limit. \ If a
point-charge mode of the current loop is used, then one becomes aware that the
magnetic moment magnitude is $m=qu_{q}R/\left(  2c\right)  =q\omega
R^{2}/\left(  2c\right)  $, and the limit of a very small radius means that
the current $I=q\omega R/\left(  2\pi R\right)  =q\omega/\left(  2\pi\right)
$ must diverge. \ If speed of the charges is less than $c,$ then the charge
density must diverge.

Use of the \textquotedblleft ideal\textquotedblright\ magnetic moment model
also tends to ignore the possibility of Faraday induction because the area
$\pi R^{2}$ of the loop is taken as negligible. \ Thus, the Faraday induction
due to the changing magnetic field of a passing charge $e$ tends to be ignored
since the area of the \textquotedblleft ideal\textquotedblright\ magnetic
moment is so small. \ At the same time, the magnetic force on the
\textquotedblleft ideal\textquotedblright\ magnetic moment due to a passing
charge $e$ moving in the same plane as the magnetic dipole is given by
$\mathbf{F}_{on\text{ }\mathbf{m}}=-\nabla\left(  -\mathbf{m\cdot B}%
_{e}\left(  \mathbf{r,}t\right)  \right)  =m\nabla B_{z}\left(  \mathbf{r,}%
t\right)  $ and, in general, will have a force component parallel to the
relative velocity between the passing charge and the magnetic moment.
\ However, this force on the\textquotedblleft ideal\textquotedblright%
\ magnetic moment seems unconnected with a force on the passing charge or with
ideas of changing electromagnetic energies. \ The use of \textquotedblleft
ideal\textquotedblright\ magnetic moments and \textquotedblleft
ideal\textquotedblright\ magnets has obscured the classical electromagnetic
interaction between a magnet and a passing charge.

\subsection{Effect First Order in the Relative Velocity Between the Inertial
Frames}

\subsubsection{Relativistic Effects of Order $V^{2}/c^{2}$}

Most relativistic experimental effects involve order $\left(  V/c\right)
^{2}$ where $V$ is the relative velocity between the inertial frames. \ Thus,
the Michelson-Morley experiment involves length contraction between inertial
frames which is second order in the relative velocity between the frames.
\ Similarly, the slowing down of decays for moving unstable particles involves
time dilation which is again second order. \ In contrast, the discrepancy
between the results from the point-charge model and the \textquotedblleft
ideal\textquotedblright\ magnetic moment model corresponds to a first-order
effect in the relative velocity $V$ between the frames. \ 

\subsubsection{Relativistic Effects of Order $V/c$}

In the present point-charge model for a magnetic moment, there is a physical
effect which is first order in the relative velocity $V$ between the inertial
frames. \ If we consider an external charge $e$ passing our current loop, the
existence of a force on the charge $e$ in the direction of the velocity
depends upon describing the system in terms of a \textit{point-charge model
for the current loop}. \ Using the point-charge model, there will be forces
upon both the charge $e$ and upon the current loop leading to a relative lag
or lead depending upon which side of the loop the charge passes. \ In the $S$
inertial frame in which the current loop has no average velocity, the forces
are associated with magnetic fields of the charge $e$ creating a magnetic
force on the current loop and a Faraday induction effect of order
$Vu_{x}/c^{2}$ leading to a force back on the charge $e$. \ In the frame
$S^{\prime}$ in which the current loop is moving and the charge $e$ is
initially at rest, all the forces are electric attractions or repulsions of
order $Vu_{x}/c^{2}$. \ This situation is analyzed in detail in the
literature.\cite{B2023} \ However, if one goes to the \textquotedblleft
ideal\textquotedblright\ magnetic moment model for the magnetic moment, then
any classical electromagnetic analysis is obscure at best. \ Thus the Faraday
induction becomes problematic in the $S$ inertial frame because the size of
the current loop is neglected; also, in the $S^{\prime}$ inertial frame, there
is no electric field in the direction of the relative velocity of the
\textquotedblleft ideal\textquotedblright\ magnetic moment.

\subsubsection{The Fizeau Experiment\ }

The Fizeau experiment\cite{Fizeau} involving light traveling in moving water
is one of the few natural phenomena involving an effect first order in the
relative velocity between the inertial frames. \ The effect is described in
terms of relativistic addition of velocities where at a single spacetime point
the transformation of the velocity between $S$ and $S^{\prime}$ inertial
frames is%

\begin{equation}
u_{x}^{\prime}=\frac{u_{x}-V}{1-(Vu_{x}/c^{2})}\approx u_{x}\left(
1+\frac{Vu_{x}}{c^{2}}\right)  -V.
\end{equation}
The \textit{nonrelativistic} addition of velocities gives simply
$u_{x}^{\prime}=u_{x}-V$, but the additional relativistic correction in order
$Vu_{x}/c^{2}$ is needed to bring high speeds down to values less than $c$ in
any inertial frame. \ In the Fizeau experiment, the water provides the
relative motion between the inertial frames of water and lab, and the speed of
light $c/n$ relative to the water (where $n$ is the index of refraction of the
water) provides the second velocity. \ 

\subsubsection{The Interaction of a Magnet and a Passing Charge}

A second example of a first-order effect (made famous by the claims of
Aharonov and Bohm)\cite{AB1959} involves the interaction of a passing charge
$e$ and a magnet. \ In the case of a magnet and passing charge, the relative
velocity $V$ between the charge and the magnet provides the relative velocity
between the inertial frames $S$ and $S^{\prime}$, and the speed $u_{x}$ of the
point charges in the magnet provides the second velocity. \ This situation, of
course, is directly related to the calculations in the present article where
the speed of the point charge $+q$ in the $S^{\prime}$ inertial frame involves
exactly the relativistic correction term $Vu_{x}/c^{2}$. \ In the $S^{\prime}$
inertial frame, the motion of the charge $+q$ has a larger
\textit{relativistic} slowing-down factor when it is moving at high speed
(nonrelativistically $\omega R+V$) than when it is moving with the slower
speed (nonrelativistically $\omega R-V$). \ Thus the positive point-charge
model for the current loop develops a relativistic electric dipole moving
where the relativistic $1/c^{2}$ correction to high-speed motion is on the
positive side of the dipole and the slower-speed relativistic correction is on
the negative side of the dipole. \ This situation involving
\textit{relativistic} speed corrections is just the reverse of the
$V^{2}/c^{2}$ change in the density of a of a line charge $\lambda$\ of finite
length which is Lorentz contracted in the direction of motion, but the total
charge is Lorentz invariant. Thus, for the line charge, the faster the line
charge moves relative to some inertial frame, the larger the charge density of
the finite line charge. \ 

\subsection{Conclusion}

Once again, we emphasize that tensor transformations of mathematical
representations for electromagnetic quantities hold only at a single spacetime
point. \ Here we have given an example showing that \textit{averages} over an
extended point-charge model of a current loop do not give reliable answers
under Lorentz transformations between inertial frames. \ We also emphasize
that \textit{continuous} current distributions can disguise the point-charge
nature of the Lorentz transformations for the currents at a spacetime point as
seen in different inertial frames. \ Finally, we point out that the use of an
\textquotedblleft ideal\textquotedblright\ magnetic dipole \ or an
\textquotedblleft ideal\textquotedblright\ magnet, which appear in the
literature, disguises appropriate Lorentz transformations between inertial
frames. 

\section{Acknowledgement}

I wish to thank Professor V. Parameswaran Nair for a helpful discussion.

\section{Data and Conflicts}

There is no new data associated with this article.

The author is not aware of any conflicts of interest.\

TensorTrans15.tex \ \ \ March 31, 2024
\end{document}